\documentclass[%
aip,
amsmath,amssymb,
reprint,%
]{revtex4-1}

\usepackage{graphicx}
\usepackage{dcolumn}
\usepackage{bm}
\usepackage{siunitx}			

\usepackage[utf8]{inputenc}
\usepackage[T1]{fontenc}
\usepackage{mathptmx}
\usepackage{xcolor}
\usepackage{etoolbox}
\usepackage{booktabs}
\usepackage{stmaryrd}
\usepackage{float}

\DeclareSIUnit{\nW}{nW}	
\DeclareSIUnit{\pW}{pW}	
\DeclareSIUnit{\THz}{THz}	
\DeclareSIUnit{\fm}{fm}	
\DeclareSIUnit{\pm}{pm}	

\begin{document}
	
	\preprint{AIP/123-QED}
	
	\title[SOI-based micro-mechanical terahertz detector operating at room-temperature]{SOI-based micro-mechanical terahertz detector operating at room-temperature}
	
	\author{K. Froberger}
		\affiliation{IEMN, Institue of Electronics, Microelectronics and Nanotechnology,  CNRS UMR 8520, Université de Lille, 59652 Villeneuve d'Ascq, France}
	\author{B. Walter}
		\affiliation{Vmicro SAS, Villeneuve d'Ascq, France}
	\author{M. Lavancier}
		\affiliation{IEMN, Institue of Electronics, Microelectronics and Nanotechnology,  CNRS UMR 8520, Université de Lille, 59652 Villeneuve d'Ascq, France}
    \author{R. Peretti}
		\affiliation{IEMN, Institue of Electronics, Microelectronics and Nanotechnology,  CNRS UMR 8520, Université de Lille, 59652 Villeneuve d'Ascq, France}
	\author{G. Ducournau}
		\affiliation{IEMN, Institue of Electronics, Microelectronics and Nanotechnology,  CNRS UMR 8520, Université de Lille, 59652 Villeneuve d'Ascq, France}
    \author{J-F. Lampin}
		\affiliation{IEMN, Institue of Electronics, Microelectronics and Nanotechnology,  CNRS UMR 8520, Université de Lille, 59652 Villeneuve d'Ascq, France}
	\author{M. Faucher}
		\affiliation{IEMN, Institue of Electronics, Microelectronics and Nanotechnology,  CNRS UMR 8520, Université de Lille, 59652 Villeneuve d'Ascq, France}
	\author{S. Barbieri}
		\email{stefano.barbieri@iemn.fr}
		\affiliation{IEMN, Institue of Electronics, Microelectronics and Nanotechnology,  CNRS UMR 8520, Université de Lille, 59652 Villeneuve d'Ascq, France}
	\date{\today}

	\begin{abstract}
		We present a micro-mechanical terahertz (THz) detector fabricated on a silicon on insulator (SOI) substrate and operating at room-temperature. The device is based on a U-shaped cantilever of micrometric size, on top of which two aluminum half-wave dipole antennas are deposited. This produces an absorption extending over the $\sim 2-3.5$THz frequency range. Due to the different thermal expansion coefficients of silicon and aluminum, the absorbed radiation induces a deformation of the cantilever, which is read out optically using a \SI{1.5}{\um} laser diode. By illuminating the detector with an  amplitude modulated, 2.5 THz quantum cascade laser, we obtain, at room-temperature and atmospheric pressure, a responsivity of \SI{\sim 1.5e8}{\pm\per\W} for the fundamental mechanical bending mode of the cantilever. This yields an noise-equivalent-power of \SI{20}{\nW\per\sqrt{\Hz}} at 2.5THz. Finally, the low mechanical quality factor of the mode grants a broad frequency response of approximately 150kHz bandwidth, with a response time of \SI{\sim 2.5}{\us}.
	\end{abstract}
	
	\maketitle

	
		In the THz range, the fastest room-temperature commercial  detectors are Schottky diodes, providing bandwidths of several tens of GHz and noise-equivalent-powers (NEPs) \si{< 100 pW/Hz^{1/2}}  up to \si{\sim 2.5 \THz}. \cite{Virginia} At higher \si{\THz} frequencies the only available options are pyroelectric detectors or thermopiles, featuring NEPs of \si{ 1-10 nW/Hz^{1/2}}, and response times of \si{\sim 10-100 ms}. \cite{Dexter,Sizov2018} The commercial availability of cheap, sensitive and faster uncooled detectors operating above 2THz, reaching bandwidths of a few 100\si{\kHz} or beyond, would be a turning point for a number of applications, including THz near-field imaging, THz spectroscopy and sensing, etc. To this end a technology that is actively investigated and developed is that of silicon-based micro-electro-mechanical systems (MEMS) bolometers. In this family of detectors we have, on one side, classical thermal micro-bolometers, where the absorbed THz radiation changes the temperature-dependent electrical conductivity of e.g. a suspended silicon (Si) membrane. Such bolometers have been integrated in CMOS compatible multipixel arrays, with reported NEPs of \si{\sim 10pW/Hz^{1/2}} and response times in the \si{\sim 10ms} range. \cite{Simoens2014,Sizov2018} Another type of MEMS bolometers exploits the so-called bi-material or bi-layer effect. Here, the temperature increase induced by the absorbed electromagnetic radiation produces a mechanical deflection due to the mismatch between the coefficients of thermal expansion (CTEs) of two materials forming a suspended micro-mirror/micro-cantilever. Such mechanical deflection is then read-out optically or electrically. These MEMS thermal detectors,  typically based on metallized silicon nitride or silicon oxide, were first developed in the mid-infrared range, and their first demonstration using a thermomechanical micro-cantilever dates back to \num{1996}. \cite{Datskos1996} Since then, several groups have demonstrated focal plane array (FPA), real-time	imaging cameras, first with optical readout, and later with a capacitive readout. \cite{Manalis1997,Perazzo1999,Grbovic2006,Hunter2006,Steffanson2014} Owing to the simplicity of sensor’s micro-fabrication and the possibility of using standard optical components, many research groups have focused on optical readout. Indeed, this method is contact-less and does not require on-chip electronics and complex wiring architectures, significantly lowering the manufacturing costs. 
		
		In \num{2008}, the first THz imaging based on a thermal MEMS array was demonstrated. \cite{Grbovic2008} In bi-material THz thermal MEMS detectors the absorbing element is a sheet of semiconductor, e.g. (\si{SiO_2}), of a few hundreds \si{\um} side, covered by a metal layer or a 2D metallic metamaterial absorber. \cite{Grbovic2008,Alves2013,Alves2018} The size of the radiation absorber and its thermal isolation from the structural anchor are appropriately designed to maximize deflection, while maintaining a response time in the 10ms range to cope with video imaging requirements. Indeed, so far, bi-material THz thermal MEMS have been almost exclusively conceived and developed to realize FPAs, with reported NEPs in the \si{10 \nW\per\sqrt{\Hz}} range. \cite{Alves2013} On the other hand, as pointed out above, much faster single pixel THz detectors are needed in many real-life applications. \cite{Hirakawa2021} To this end a considerable increase in the response speed can be obtained by reducing the size of the thermomechanical elements: for example, the thermal time constant of a simple cantilever is proportional to the square of its length. However, reducing the structure geometrical dimensions below the wavelength of the incident THz radiation, hinders efficient radiation absorption. An elegant solution to this problem is the use of metallic antennas as absorbing element. Recently, the coupling of free space radiation at \SI{2.5}{\THz} ($\lambda$~=~\SI{120}{\um}) to a suspended GaAs cantilever of sub-wavelength length (around \SI{15}{\um}) was demonstrated by embedding the latter in a Au split-ring resonator antenna. \cite{Belacel2017} 
		
		In this work we exploit this concept to demonstrate a MEMS THz bolometer based on a \si{\um}-size bi-layer microcantilever structure, yielding a thermal response time of 2-3 \si{\us}. Two aluminum (Al) half-dipole antennas are used as absorbing elements, allowing a broad spectral response, from $\sim$2 to $\sim$3.5THz, and an estimated close to unity collection efficiency of the incident radiation. The device is developed entirely on a SOI-photonic substrate, i.e. with a processing compatible with CMOS microelectronics fabrication standards and potentially allowing the addition of on-chip passive photonics structures, suitable to implement an integrated readout. \cite{Anetsberger2009,Kim2013} 
	
\begin{figure}
	\includegraphics[width=0.47\textwidth]{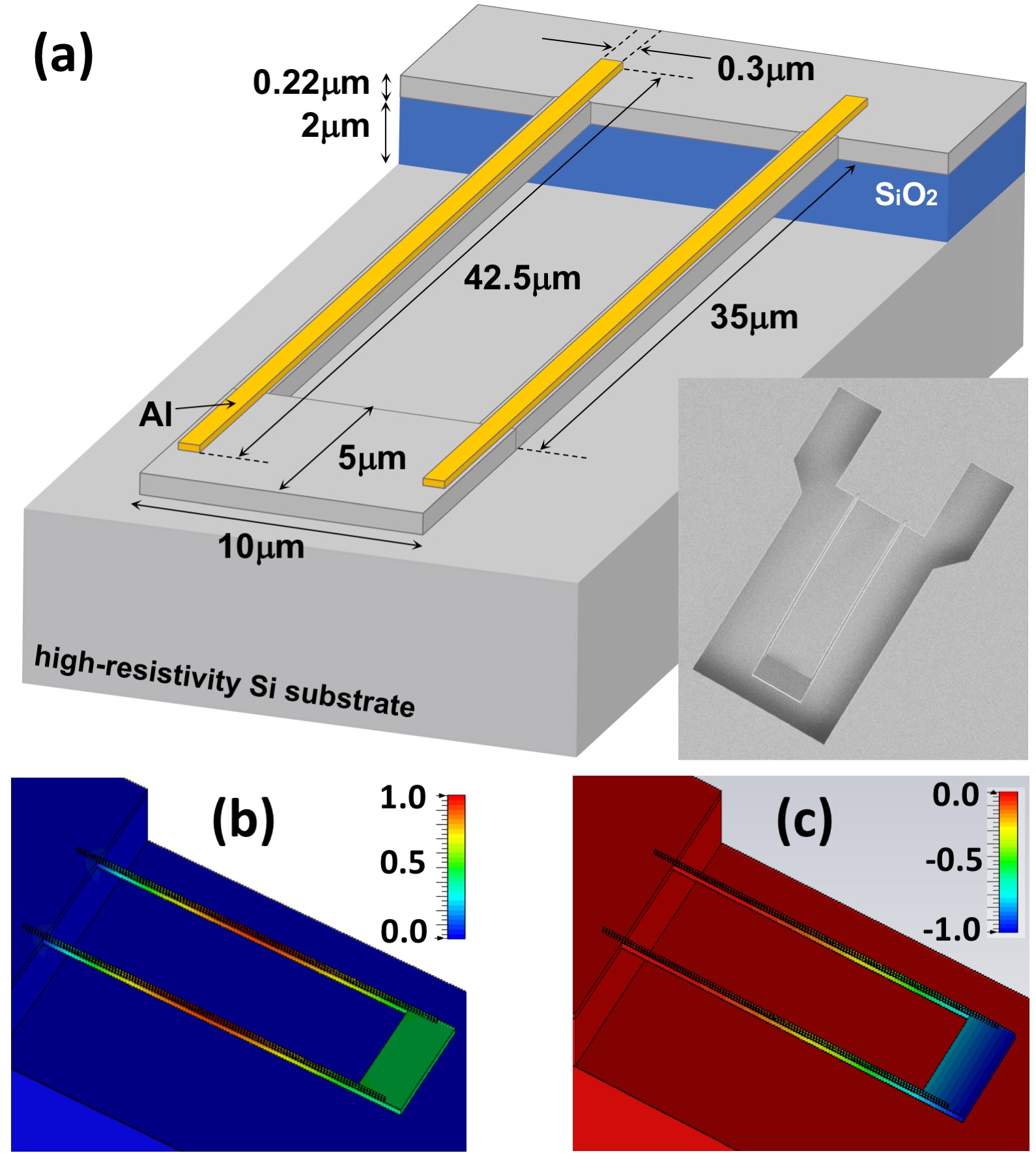}
	\caption{\label{fig:Design} (a) Schematic of the cantilever geometry with all relevant dimensions. The Al antennas are 80nm thick. The inset shows a SEM image of the fabricated device. (b) Temperature profile (in normalised units) of the cantilever simulated using CST Microwave Studio\textsuperscript{\textregistered}. The cantilever is illuminated by a 3THz plane wave of constant amplitude. (c) Simulated displacement (in normalised units) of the cantilever in the direction perpendicular to the substrate (out-of-plane), induced by the bi-layer effect. }
	\end{figure}

    The detector was fabricated with e-beam lithography and its  geometry is shown schematically in Fig.\ref{fig:Design}(a).  \si{SiO_{2}} is used as sacrificial layer to obtain a suspended, U-shaped Si structure, consisting of two \SI{35}{\um}-long cantilevers, \SI{500}{\nm}-wide and \SI{220}nm-thick, connected at one end by a \SI{5}{\um}$\times$\SI{10}{\um} rectangular pad, used for the optical readout (see below). Two \SI{80}{\nm}-thick, \SI{42.5}{\um}-long and \SI{300}{\nm}-wide Al dipole antennas, designed to resonate close to 3THz, are deposited on top of the cantilever’s arms. Absorption of the impinging THz radiation induces an \textit{ac} current with a maximum at the center of the antennas, giving rise, through joule heating, to the temperature profile shown in Fig.\ref{fig:Design}(b), obtained through FDTD simulations with the structure illuminated by a plane wave at 3THz. As a result, due to the fact that Al and Si have different CTEs, the structure bends in the direction perpendicular to the high-resistivity Si-substrate (out-of-plane movement). The resulting simulated displacement in arbitrary units in shown in  Fig.\ref{fig:Design}(c). The choice of Al, is dictated by the need to maximize the difference in CTEs compared to Si in order to amplify the bi-layer effect. In this respect, Al is among the best metals commonly used in MEMS, with a CTE of around \si{25 \times 10^{-6} K^{-1}}, i.e. 10 times larger than that of Si.

\begin{figure}
\includegraphics[width=\linewidth]{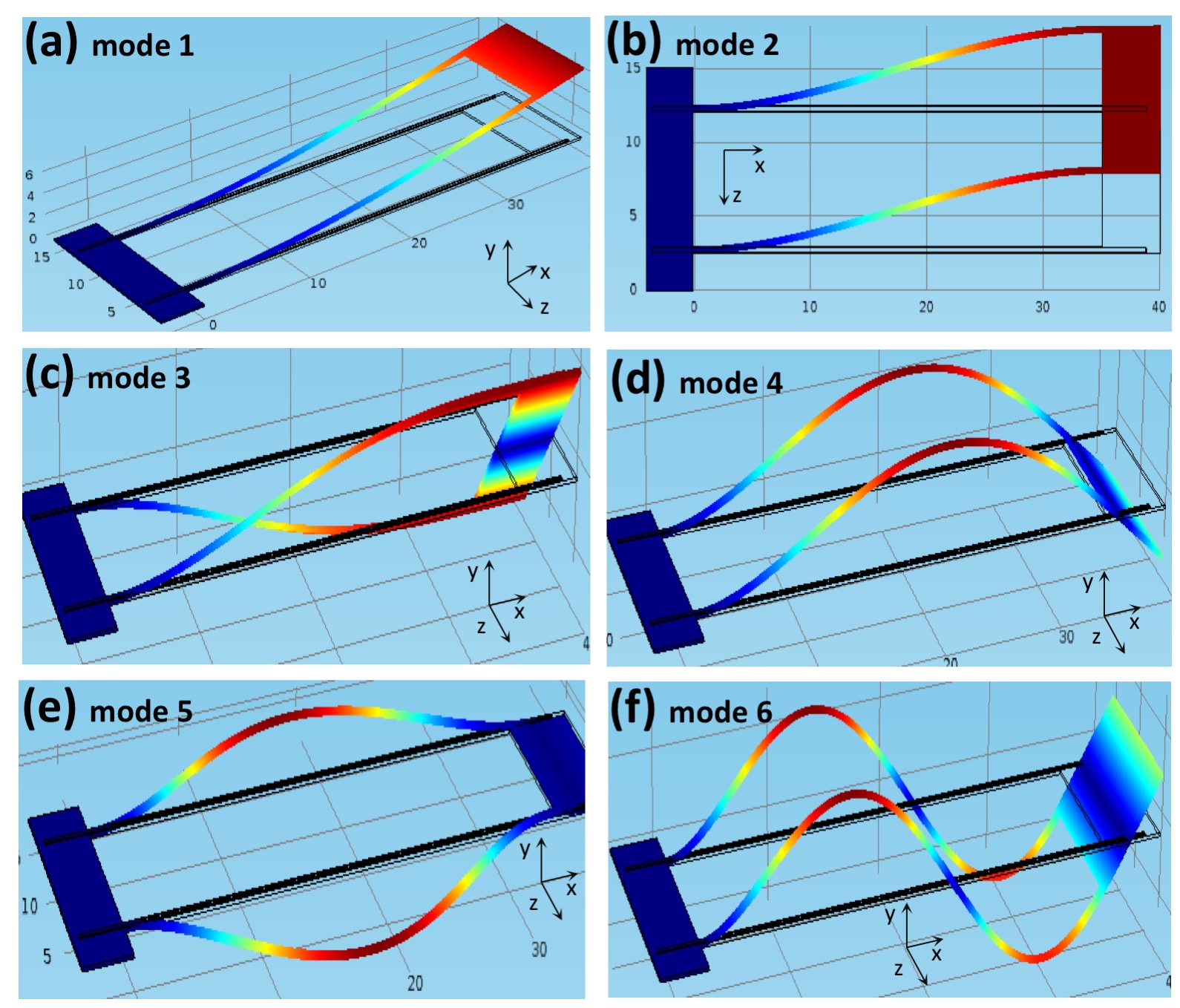}
\caption{Profile of the first 6 mechanical modes of the micro-bolometer simulated on Comsol\textsuperscript{\textregistered}. (a) Mode 1: out-of-plane bending mode at \SI{101}{\kHz}. (b) Mode 2: in-plane mode at \SI{389}{\kHz}. (c) Mode 3: torsional mode at \SI{639}{\kHz}. (d) Mode 4: out-of-plane second bending mode at \SI{960}{\kHz}. (e) Mode 5: in-plane mode at \SI{1707}{\kHz}. (f) Mode 6: out-of-plane third bending mode at \SI{2940}{\kHz}.}
\label{fig:mecanique}
\end{figure}

		As we shall see below, the sensitivity of the MEMS detector can be enhanced through the resonant excitation of mechanical modes. Using Comsol \textsuperscript{\textregistered}, we simulated the shape and the frequency of the first 6 mechanical modes of the cantilever up to 3.5 MHz. As shown in Fig.\ref{fig:mecanique}, modes 1 (Fig.~\ref{fig:mecanique}(a)), 4 (Fig.~\ref{fig:mecanique}(d)) and 6 (Fig.~\ref{fig:mecanique}(f)) are pure out of plane bending modes. These are the modes of interest, as they will be excited by the bilayer effect induced by an incident, amplitude modulated THz wave (see Fig.1(b),(c)). On the contrary, we expect that modes 2 (Fig.~\ref{fig:mecanique}(b)) and 5 (Fig.~\ref{fig:mecanique}(e)), corresponding to in-plane vibrations, and mode 3 (Fig.~\ref{fig:mecanique}(c)), a twisting mode, will be excited much less efficiently.
		
		Let's consider an incident THz beam, with a power modulated at a frequency $\omega$: $P_{THz} = P^{0}_{THz}(1+cos(\omega t))/2$. The photo-thermal responsivity of the device (in units of m/W) associated to the \textit{n}-th mechanical mode of the cantilever can be written as: \cite{Belacel2017,Metzger2008}
		
		\begin{equation}
			R_{n}^{ph}(\omega) = \frac{|z(\omega)|}{P^{0}_{THz}} = F_{ph}\times |Y(\omega)H_{n}(\omega)|,
			\label{eq:Responsivity}
		\end{equation}
			with
		\begin{equation}
			Y(\omega) = \frac{1}{1+i\omega\tau},
			\label{eq:heat propagation}
		\end{equation}
		\begin{equation}
			H_{n}(\omega) = \frac{1/m_{eff,n}}{\omega_{n}^{2} - \omega^{2}+i\omega_{n}\omega/Q_{n}}.
			\label{eq:MecOscTransfer}
		\end{equation}	
		
		In Eq. \ref{eq:Responsivity}, $n$ is the mode number, $z(\omega)$ is the cantilever out-of-plane displacement, and $F_{ph}$ is the photo-thermal force per unit incident power, $P^{0}_{THz}$, induced by the amplitude modulated THz wave. $Y(\omega)$ takes into account the process of heat diffusion through the cantilever, and behaves as a 1st order low-pass filter with a characteristic time constant $\tau$. Finally, $H_{n}(\omega)$ is the transfer function of the mechanical mode with resonant frequency $\omega_{n}$ and effective mass $m_{eff,n}$. Since the  structure cannot be considered as a simple beam, it is not trivial to theoretically derive $\tau$. However, from a transient simulation using CST Microwave Studio\textsuperscript{\textregistered}, the time needed to reach steady state is found to be of approximately \SI{15}{\us}.
		
		Under ideal operation, the noise floor of the THz opto-mechanical detector is limited by the thermal, or brownian motion, of the cantilever at the operating temperature $T$. The noise power spectral density (PSD) of the $n^{th}$ mechanical mode (in units of $m^{2}$/Hz) is given by: \cite{Hauer2013}
			
			\begin{equation}
				S_{zz}(\omega) = \frac{4k_BT\omega_n m_{eff,n}}{Q_n}
				|H_{n}(\omega)|^2,
				\label{eq:PSDModel}
			\end{equation}
			where $k_B$ is the Boltzmann constant. From Eqs.\ref{eq:Responsivity}-\ref{eq:PSDModel}, we finally obtain the expression of the NEP:
		
		    \begin{equation}
				P_{NEP}(\omega)=\frac{S_{zz}(\omega)^{1/2}}{R_{n}^{ph}(\omega)}= \frac{2}{F_{ph}|Y(\omega)|}\sqrt{\frac{k_BTm_{eff}\omega_n}{Q_n}}
				.
				\label{eq:NEP}
			\end{equation}	
			
		From Eqs.\ref{eq:Responsivity} and \ref{eq:NEP} we find that exploiting a cantilever with a high mechanical Q-factor leads to an increase of responsivity and a reduction of $P_{NEP}$. At the same time, a high Q-factor will entail, through the mechanical mode transfer function (Eq.\ref{eq:MecOscTransfer}), a detector's frequency response with a narrow band centered around $\omega_n$, generally resulting in significant signal distortion (of e.g. a THz wave with an arbitrarily time-varying envelope), which can be unpractical for applications. \cite{Todorov2021} To mitigate this problem, in this work we have deliberately chosen to work with a cantilever with a low Q-factor, leading to a broader and less peaked frequency response (see below). 
		
\begin{figure}
				\includegraphics[width=\linewidth]{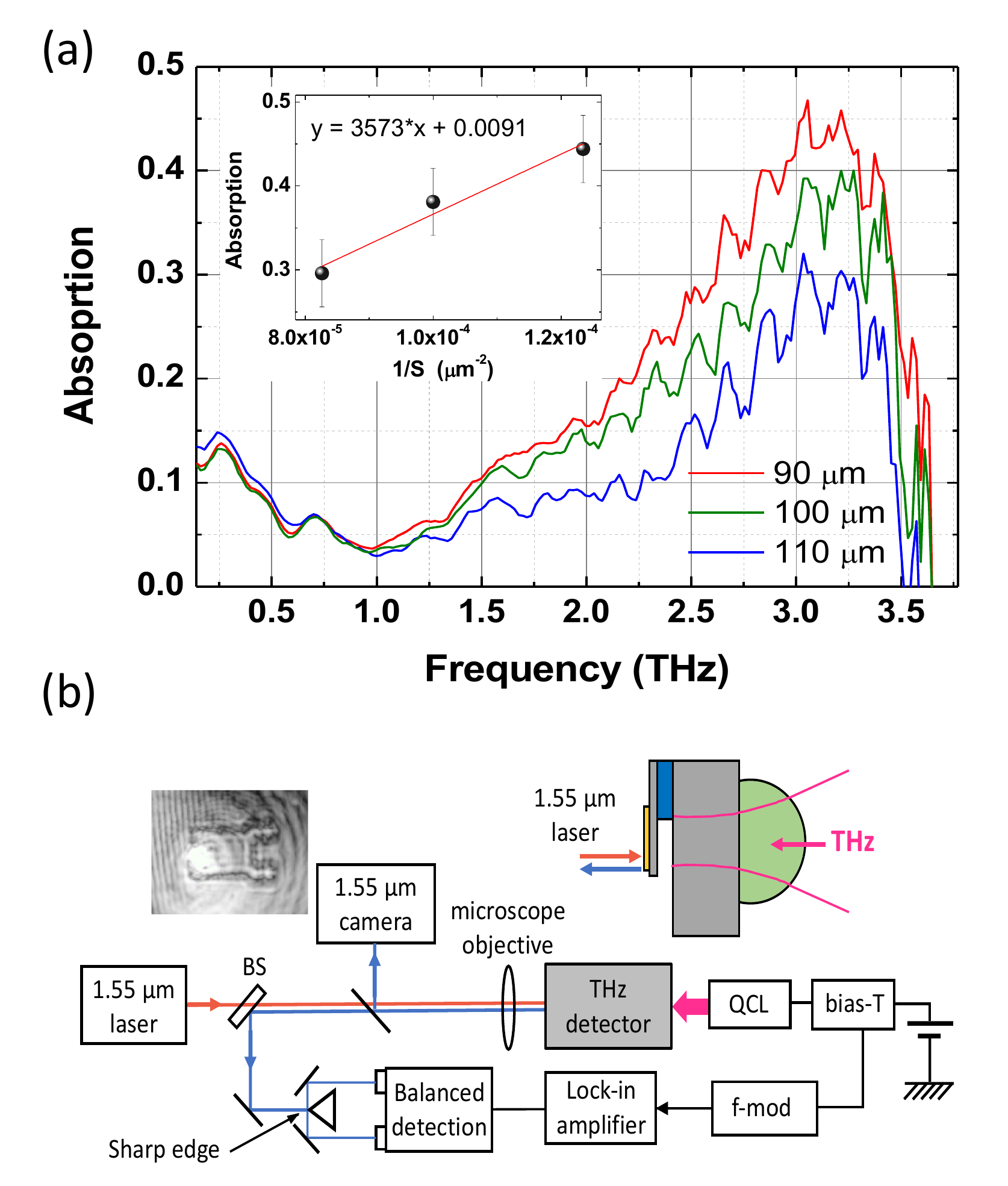}
				\caption{(a) THz TDS absorption spectra of sets of square matrices obtained by repeating periodically the detector element of Fig. \ref{fig:Design}(a) with periods $p= 90$ (red), $100$ (green)  and \SI{110}{\um} (blue). Inset: peak absorption $vs$ inverse unit cell area ($S = p^{2}$). The slope of the fit yields an antenna effective area of \SI{3750}{\um^2}. (b) Experimental setup for the optical readout of the cantilever's mechanical movement. The beam from at \SI{1550}nm diode laser is focused on the cantilever's end pad. The reflected beam is used to both visualize the device on an InGaAs camera (an image example is shown in the left inset) and measure the cantilever's deflection with a balanced photodetector. For THz excitation, the current of a QCL is modulated sinusoidally, generating an amplitude modulated beam that is focused on the device through a Si hemispherical lens (in green in the right inset) positioned on the Si substrate.}
				\label{fig:Setup}
			\end{figure}

		
		A commercial THz time domain spectroscopy (TDS) system in transmission geometry was used to quantify the absorption of the double-dipole antenna. To this end, sets of square matrices of 2.5mm side were fabricated by repeating periodically the detector element of Fig.\ref{fig:Design}(a) with periods $p= 90, 100$ and $110 \mu$m.  
		Transmission spectra ($T(\nu)$) were obtained from the ratio between the spectrum of the THz pulse transmitted through the matrix and that of a reference pulse passing trough the unprocessed SOI substrate (to avoid Fabry-Perot fringes due to multiple reflections, only the first transmitted THz pulse was considered in the analysis - see \textcolor{blue}{Supplementary document}). The corresponding absorption (1-$T(\nu)$) spectra are shown in Fig.\ref{fig:Setup}(a).
		We observe a broad absorption peak, of approximately 1 THz FWHM, with a maximum close to 3THz. The peak asymmetry is an artifact produced by the limited spectral bandwidth of the TDS system, resulting in a fast loss of signal-to-noise ratio above $\sim 3.5$ THz (see \textcolor{blue}{Supplementary document}). 
		With increasing period, we find a monotonic decrease of the absorption as a result of the detector's antennas being progressively farther apart. This occurs without any substantial change in the spectral shape, demonstrating a negligible coupling between neighboring double-dipole antennas. Therefore, we can assume that the fraction of THz power absorbed by the antenna is roughly given by the ratio between the antenna effective area ($A_{eff}$) and the surface ($S = p^{2}$) of the unit cell of the matrix of detectors. In the inset of Fig.\ref{fig:Setup}(a) the peak absorption is plotted as a function of $1/S$. From the slope of the linear fit we obtain $A_{eff} \sim$3700$\mu$m$^2$. A rough estimate of $A_{eff}$ is given by sum of the effective areas of two dipole antennas, yielding $2\times0.13\lambda^2 = 2600\mu$m$^2$ at 3THz.\cite{Orfanidis}
		
		In the remaining of this work, the performance of our MEMS bolometer as single element detector is characterised using, as THz source, a quantum cascade laser (QCL) emitting at \SI{2.5}{\THz}.
		To this end we have setup the experiment reported schematically in Fig.~\ref{fig:Setup}(b), allowing the optical readout of the cantilever's movement under THz QCL excitation. \cite{Belacel2017} For the optical readout we focused, using a microscope objective, a low intensity-noise, single mode \SI{1.55}{\um} diode laser (RIO-Orion) on the rectangular Si pad at the end of the cantilever (Fig.~\ref{fig:Design}(a)). The incident optical power was of a few mW. We chose to work at \SI{1.55}{\um} rather than 800nm to minimise absorption in Si.  Any vibration of the cantilever deflects the reflected the laser beam. As shown in Fig.~\ref{fig:Setup}(b) the deflection is probed using a sharp-edge prism followed by a balanced detection based on two InGaAs photodiodes and connected to a lock-in amplifier (Zurich Instruments UHFLI). Reducing the excess laser intensity noise by equalizing the powers on the photodiodes, allows reaching the shot noise limit at high frequency ($\gtrsim 4$MHz), while at lower frequencies the noise floor is given by the brownian motion of the cantilever (see below).

\begin{figure}
				\includegraphics[width=\linewidth]{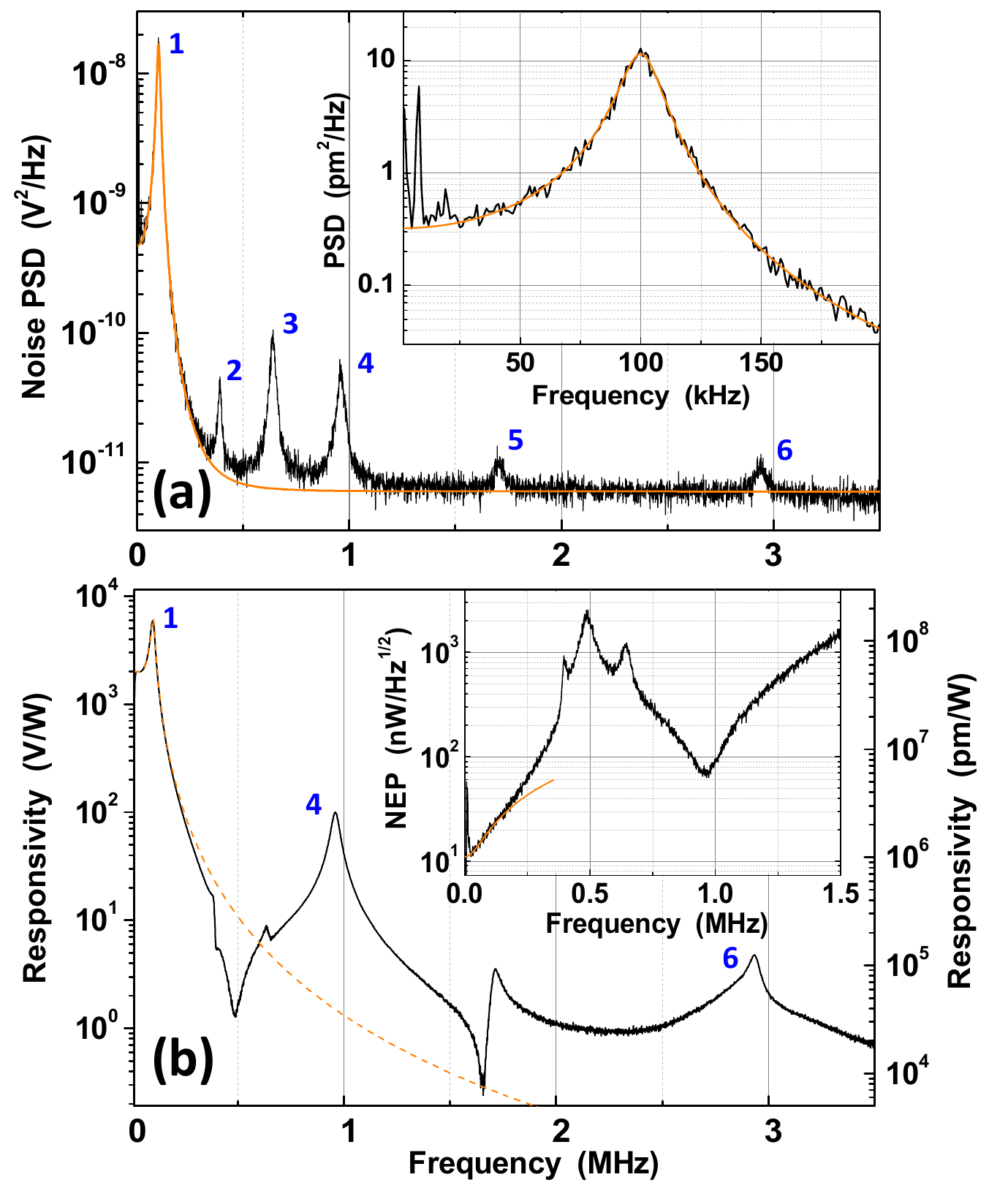}
				\caption{(a) Noise PSD. Modes 1, 4 and 6 correspond to pure out-of-plane oscillations (see Fig.\ref{fig:mecanique}). The orange solid-line is the result of a best fit of the fundamental mode of the cantilever using Eq.\ref{eq:PSDModel} (see main text).  (b) Responsivity of the device at \SI{2.5}{\THz}. Only modes 1, 4 and 6 are excited efficiently. The orange dashed line is a best fit of the fundamental mode obtained through Eqs.\ref{eq:Responsivity}-\ref{eq:MecOscTransfer}. Inset. Experimental $P_{NEP}$ $vs$ modulation frequency. The solid orange line is given by (11 nW/Hz$^{1/2}$) $\times\sqrt{1+(2\pi f \tau)^2}$, with $\tau = 2.4 \mu$s.}
				\label{fig:NoiseResp}
			\end{figure}

		The measured thermal noise PSD of the MEMS detector (QCL switched off) is displayed in Fig.~\ref{fig:NoiseResp}(a) in units of V$^2$/Hz, as a function of frequency $f = \omega/2\pi$. We find 6 peaks: within $\sim 1\%$ their frequencies are equal to those of the simulated mechanical modes (see caption of Fig.~\ref{fig:mecanique}). As discussed previously, modes n.1, 4 and 6 correspond to pure out-of-plane oscillations, while mode n.3 is a twisting mode. As such they all change the reflection angle of the readout laser beam. Instead modes n.2 and 5 are associated to a purely in-plane movement, which, however, can still produce a weak but measurable power change on the photodiodes because the laser spot is located close to one edge of the cantilever end pad (see the left inset of Fig.~\ref{fig:Design}(b)). This was done on purpose to allow the detection of these modes and validate the modelling. 
			
		The orange solid-line in Fig.~\ref{fig:NoiseResp}(a) is the result of a fit of the measured thermal noise  PSD of the fundamental mode of the cantilever, obtained using Eq.\ref{eq:PSDModel} ($\omega=f/2\pi$) with the addition of a constant term to take into account the white noise-floor at high frequencies. \cite{Aspelmeyer2014,Hauer2013} The other fit parameters are the mode frequency ($f_1 = \omega_1/2\pi$), the quality factor ($Q_1$), and a multiplication factor $M$. A shown in Fig.~\ref{fig:NoiseResp}(a) the fit is excellent, yielding $f_1 = 100.7$kHz and $Q_1=6$. The ratio between the fitting curve and the theoretical expression of the noise PSD of Eq.\ref{eq:PSDModel}, obtained with the computed effective mass of the fundamental mode of 34pg,\cite{Hauer2013} allows to calibrate the cantilever displacement, i.e. to determine the detection system's conversion factor from pm to Volts, and finally express the measured thermal noise PSD of the fundamental mode in units of pm$^2$/Hz, as shown in the inset of Fig.~\ref{fig:NoiseResp}(a).
		
		To measure the detector's frequency response, the QCL beam was collimated and focused on the cantilever using two parabolic mirrors and a hemispherical Si-lens, of 0.65 numerical aperture, in contact with the back of the substrate (Fig.~\ref{fig:Design}(b)). The diffraction-limited focused spot-size, of $\sim$1800$\mu$m$^2$, is well below the antenna effective area (see above), which, taking into account the reflection at the Air/Si interface, should allow a collection efficiency of $\sim 70 \%$. The QCL was driven in continuous-wave above threshold. A sinusoidal current modulation was added with the help of a bias-T. The modulated optical power was extracted from the QCL Power/Current characteristic, that was measured with a calibrated thermopile detector. 
			
\begin{figure}
				\includegraphics[width=\linewidth]{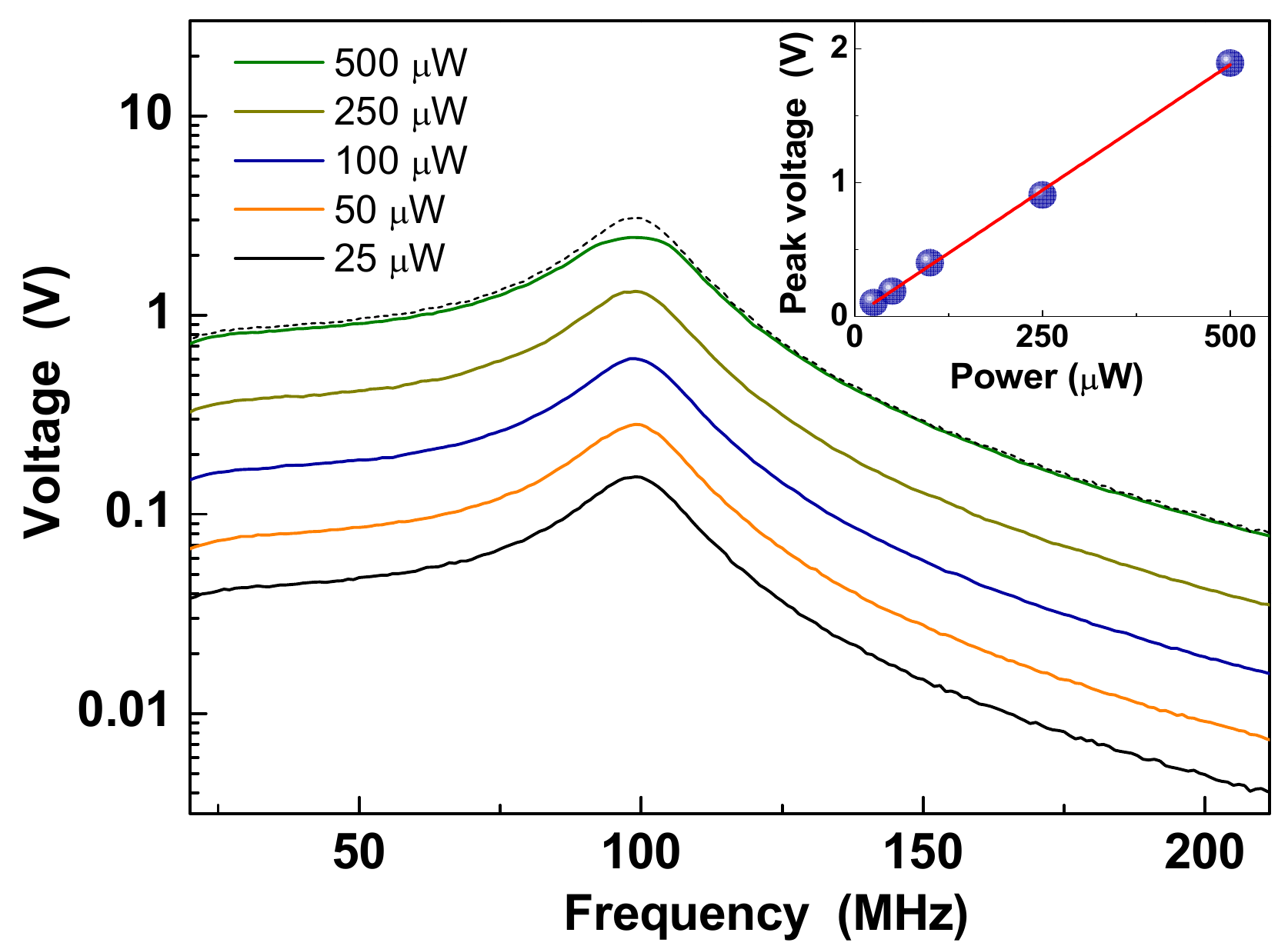}
				\caption{Lockin amplifier output voltage $vs$ modulation frequency for different THz incident powers. The dashed line is obtained by multiplying by 20 the output voltage corresponding to an incident power of $25 \mu$W (solid black line). Inset. Output voltage $vs$ power at 100kHz modulation frequency.}
				\label{fig:microsensorParameters}
			\end{figure}

		The experimental responsivity spectrum in units of V/W and pm/W (using the calibration of the thermal noise), measured with a modulated incident power of 100$\mu W$, is shown in Fig.4(b) as a function of the modulation frequency. It was obtained by sweeping the latter from \num{0} to \SI{3.5}{\MHz} and by recording the signal from the balanced photodetector with the lockin amplifier. We stress that to compute the responsivity we used the power incident on the Si lens, without any correction factor. As expected, only the out-of-plane modes are efficiently excited. The dashed orange line is a fit of the fundamental mode using Eq.\ref{eq:Responsivity}, yielding $f_1 = 100$kHz, $Q_1=5.5$ and a response time $\tau = 2.4 \mu$s. This corresponds to a time to reach steady state of 12$\mu$s (99$\%$ response), which is in good agreement with the 15$\mu$s obtained from transient simulations. We note that the relatively low Q-factor grants a a rather broad response function, with a ratio of 2.9 between the peak responsivity ($f_1 = 101$ kHz) and the value at $dc$ (see also Fig.~\ref{fig:microsensorParameters}). A flatter response could be obtained by reducing further the Q-factor.
			
	In the inset of Fig.~\ref{fig:NoiseResp}(b) we report the
	experimental NEP $vs$ modulation frequency, obtained from
	the ratio between the square root of the noise PSD and the
	responsivity (Fig.4(a),(b)). As expected from Eq. \ref{eq:NEP},
	$P_{NEP}$ follows a $\sqrt{1+(\omega\tau)^2}$ dependence at
	low frequency (solid orange line, with $\tau = 2.4 \mu$s),
	with a minimum of 10 nW/Hz$^{1/2}$ close to $dc$, and a value
	of 30 nW/Hz$^{1/2}$ at 150kHz. Another minimum at $\sim 70$
	nW/Hz$^{1/2}$ is found at the resonant frequency of mode n.4
	(2$^{nd}$ out-of-plane, $f=960$ kHz). Finally, in Fig.5(b) we
	present the evaluation of the detection linearity in the
	range $\sim 10 - 500 \mu$W. In this range the linearity is
	excellent, with the exception of a small saturation at the
	peak of the spectrum with $P^{0}_{THz} = 500\mu$W, due to a
	saturation of the lockin amplifier (compare the solid green
	and dashed curves).
	
	 In conclusion, we have designed and characterised a micro-mechanical THz detector, covering the 2-3.5THz spectral range, and fabricated on a SOI platform.  The device exploits the thermal excitation of the fundamental mechanical mode of a U-shaped Si cantilever, on top of which are deposited two half-dipole antennas resonating at 3THz. The measured $P_{NEP}$ is in the 10-30 nW/Hz$^{1/2}$ range, i.e. comparable to Ref. [15] and to that of commercially available thermopiles and pyroelectric detectors with ms response times. Compared to Ref. [15], thanks to a significantly lower mechanical quality factor (6 vs 100), we obtain a broader frequency response, extending from $dc$ to ~150kHz, with a peak at $\sim 100$kHz, and a response time of $\sim 2.5\mu$s. Reducing further the Q-factor and/or increasing the resonant mode frequency would allow an even flatter response. Future work will focus on the demonstration of detectors based on shorter antennas, for operation in the 4-7 THz range. Secondly, we plan to take advantage of the fabrication capabilities offered by the chosen SOI-photonic platform to implement an integrated photonic readout. \cite{Anetsberger2009,Kim2013} 
			
\section*{SUPPLEMENTARY MATERIAL}
See the Supplementary document for further details about the THz time domain spectroscopy measurements related to Fig.3.

	\begin{acknowledgments}
		The authors would like to thank Giorgio Santarelli for providing the balanced detection and for helpful discussions. We gratefully acknowledge Cherif Belacel for helpful discussions on the experimental setup. We acknowledge partial financial support from the French Renatech network, Nord-Pas de Calais Regional Council (grant STARS-ATENA), Fonds Européens de Développement Régional, and CPER "Photonics for Society".
	\end{acknowledgments}

\section*{AUTHOR DECLARATIONS}
\subsection*{Conflict of Interest}
The authors declare no conflicts of interest.

\subsection*{Data Availability}
The data that support the findings of this study are available from the corresponding author upon reasonable request.

	
	\section*{References}
	\nocite{*}
	\bibliography{aipsamp}
	
\end{document}